\documentclass[aps,prb,reprint,groupedaddress]{revtex4-1}
\usepackage{graphicx}

\begin{document}


\title{Antiferromagnetic ordering in organic conductor $\lambda$-(BEDT-TTF)$_2$GaCl$_4$ probed by $^{13}$C NMR}


\author{Yohei Saito$^1$}
\author{Shuhei Fukuoka$^1$}
\email{fukuoka@phys.sci.hokudai.ac.jp}
\author{Takuya Kobayashi$^1$}
\author{Atsushi Kawamoto$^1$}
\author{Hatsumi Mori$^2$}
\affiliation{$^1$Department of Physics, Hokkaido University, Sapporo 060-0810, Japan}
\affiliation{$^2$Institute for Solid State Physics, University of Tokyo, Kashiwa, Chiba 277-8581, Japan}


\begin{abstract}
The ground state of $\lambda$-(BEDT-TTF)$_2$GaCl$_4$, which has the same structure as the organic superconductor $\lambda$-(BETS)$_2$GaCl$_4$, was investigated by magnetic susceptibility and $^{13}$C NMR measurements. The temperature dependence of the magnetic susceptibility revealed an antiferromagnetic (AF) correlation with $J/k_{\rm B} \simeq$ 98 K. NMR spectrum splitting and the divergence of $1/T_1$ were observed at approximately 13 K, which is associated with the AF transition. We found that the AF structure is commensurate according to discrete NMR peak splitting, suggesting that the ground state of $\lambda$-(BEDT-TTF)$_2$GaCl$_4$ is an AF dimer-Mott insulating state. Our results suggest that the superconducting phase of $\lambda$-type salts would be located near the AF insulating phase.
\end{abstract}


\maketitle

The pairing mechanism of unconventional superconductivity is closely related to the nature of the adjacent insulating phase. In the case of organic superconductors, various mechanisms of superconductivity are suggested\cite{Schmalian1998,Merino2001}. The superconducting (SC) phase in the $\kappa$-type BEDT-TTF [BEDT-TTF(ET)=bis(ethylenedithio)tetrathiafulvalene] system is located near an antiferromagnetic (AF) Mott insulating phase, and AF spin fluctuation plays a crucial role in the emergence of the superconductivity \cite{Kanoda1997}. On the other hand, the SC phase in $\beta''$-type BEDT-TTF salts is located near the charge-ordered phase, and the relation between the charge fluctuation and the superconductivity has been discussed \cite{Merino2001,Girlando2014}.

The quasi-two-dimensional organic superconductor $\lambda$-(BETS)$_2$GaCl$_4$ [BETS = bis(ethylenedithio)tetraselenafulvalene] has attracted considerable attention. It exhibits a SC transition at approximately 5 K \cite{Kobayashi1993,Tanatar2002a}. When a magnetic field is applied exactly parallel to the conducting layers, it is suggested that a Fulde--Ferrell--Larkin--Ovchinnikov (FFLO) state is realized in the vicinity of the critical magnetic field ($H_{\rm c2}$) \cite{Coniglio2011,Uji2015}. From the heat capacity measurement, the symmetry of the SC gap is determined to be the nodal $d$-wave \cite{Imajo2016}. In $\lambda$-type salts, two donor molecules form a dimer in the conducting layers as in the $\kappa$-type BEDT-TTF system. Thus, it has been considered that the insulating phase adjacent to the SC phase in $\lambda$-type salts is also an AF Mott insulating phase and that the pairing mechanism is AF spin fluctuation. However, Tanaka \textit{et al}. investigated the magnetic nature of the insulating phase of $\lambda$-type BETS salts using the substitution effect of anion molecules. They investigated $\lambda$-(BETS)$_2$GaBr$_x$Cl$_{4-x}$ and suggested that the adjacent insulating phase is a nonmagnetic insulating state rather than the AF insulating state \cite{Tanaka1999}. However, the substitution of anions can introduce structural disorder in anion layers. Thus, further study of the insulating phase in a homogeneous system is needed.

The substitution of donor molecules also gives rise to the chemical pressure effect, and a systematic study of $\lambda$-$D_2$GaCl$_4$ ($D$= BEDT-TTF, us-BEDT-STF, and BETS [us-BEDT-STF (STF) = unsymmetrical-bis(ethylenedithio)diselenadithiafulvalene]) was performed \cite{Mori2001}. The resistivity of $\lambda$-(BEDT-TTF)$_2$GaCl$_4$ at 1.65 GPa corresponds to that of $\lambda$-(us-BEDT-STF)$_2$GaCl$_4$ at ambient pressure \cite{Mori2001}, and $\lambda$-(us-BEDT-STF)$_2$GaCl$_4$ exhibits superconductivity over 1.22 GPa \cite{Minamidate2015}.
\begin{figure}[tbp]
  \includegraphics[width=8.5cm,pagebox=cropbox,clip]{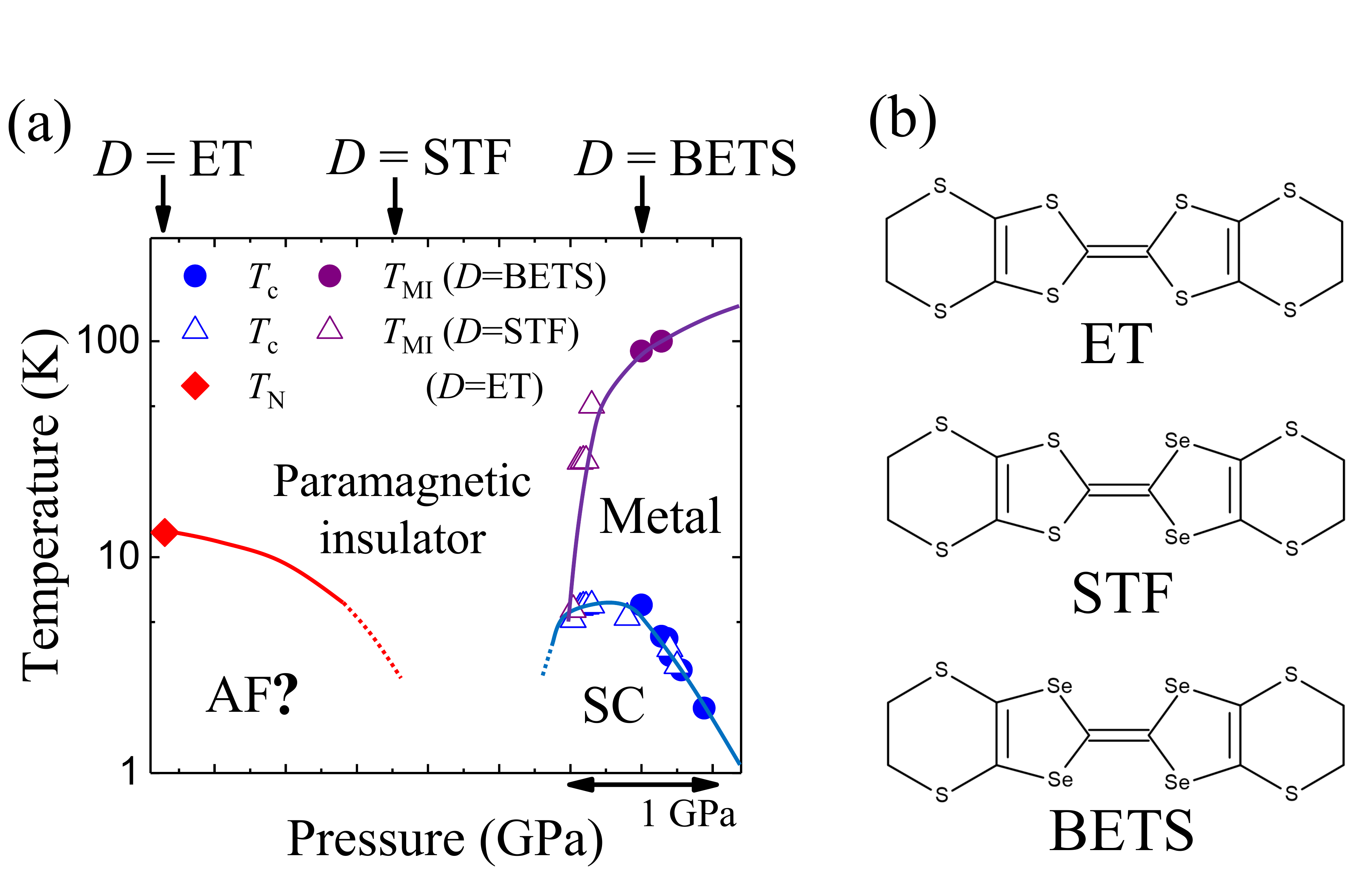}
  \caption{(Color online) (a) Pressure-temperature phase diagram of $\lambda$-$D_2$GaCl$_4$ ($D$ = ET, STF, and BETS). ET and STF correspond to BEDT-TTF and us-BEDT-STF, respectively. $T_{\rm c}$, $T_{\rm MI}$, and $T_{\rm N}$ represent the superconducting transition temperature, metal--insulator transition temperature, and N\'eel temperature, respectively. The data are cited from Refs. [10-12]. (b) Molecular structure of BEDT-TTF(ET), us-BEDT-TTF(STF), and BETS.}\label{phasediagram}
\end{figure}
These results demonstrate that $\lambda$-(BEDT-TTF)$_2$GaCl$_4$ and $\lambda$-(us-BEDT-STF)$_2$GaCl$_4$ are located in the insulating phase adjacent to the SC phase, as shown in the general phase diagram in Fig. \ref{phasediagram}(a) \cite{Kobayashi2017}. Structure of the donor molecules is shown in Fig. \ref{phasediagram}(b). Note that the SC state of $\lambda$-(BEDT-TTF)$_2$GaCl$_4$ is not confirmed, because of a lack of high-pressure experiments. Because $\lambda$-(BEDT-TTF)$_2$GaCl$_4$ does not contain structural disorder in the anion layers, a detailed study of the insulating phase is possible. At present, only an electron spin resonance (ESR) measurement in $\lambda$-(BEDT-TTF)$_2$GaCl$_4$ was performed, and the possibility of an AF transition was suggested from a decrease in the ESR intensity \cite{Mori2001}. Therefore, the magnetic nature of the insulating phase is still not clear, and it is of significant research interest.

Nuclear magnetic resonance (NMR) spectroscopy is one of the powerful techniques for investigating the magnetic nature of strongly correlated electron systems. The development of magnetic spin fluctuation can be probed by measuring the nuclear spin-lattice relaxation rate ($1/T_1$). The NMR spectrum also can detect the staggered moment of the AF phase directly. In this study, to clarify the detailed nature of the insulating ground state and determine the phase diagram of the $\lambda$-type salts, we performed $^{13}$C NMR and magnetic susceptibility measurements on $\lambda$-(BEDT-TTF)$_2$GaCl$_4$.

Single crystals of $\lambda$-(BEDT-TTF)$_2$GaCl$_4$ were prepared from the electrochemical oxidation of BEDT-TTF in 1, 1, 2-trichloroethane in the presence of TBAGaCl$_4$ \cite{Mori2001}. The magnetic susceptibility was measured using powder samples as a function of temperature from 300 to 2 K in a magnetic field of 1 T. For $^{13}$C NMR measurements, to avoid the Pake doublet problem, one of the carbons in the central C=C bond in BEDT-TTF molecules was enriched with $^{13}$C isotope by a cross-coupling method \cite{Hirose2012}. NMR experiments were performed in a magnetic field of 7 T applied in the [110] direction (almost parallel to the long axis of BEDT-TTF). The crystal orientation was determined by X-ray diffraction. The NMR spectra were obtained by the fast Fourier transformation of the spin echo signal with a $\pi/2{\rm -}\pi$ pulse sequence. The nuclear spin-lattice relaxation rate ($1/T_{1}$) was measured using the conventional saturation-recovery method.

\begin{figure}[htbp]
  \includegraphics[width=8.5cm,pagebox=cropbox,clip]{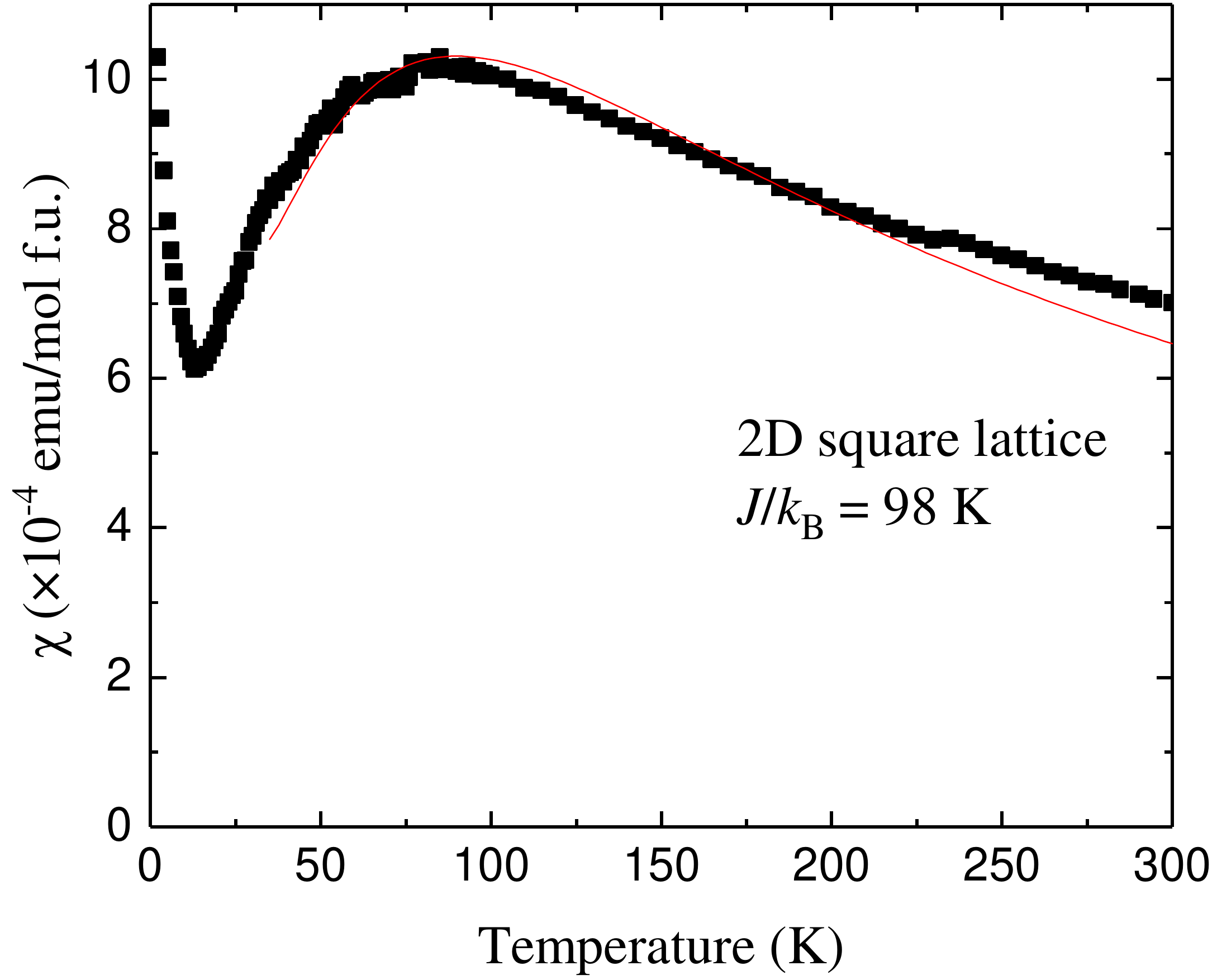}
  \caption{(Color online) Magnetic susceptibility as a function of temperature for $\lambda$-(BEDT-TTF)$_2$GaCl$_4$. The solid curve represents the result of two-dimensional $S=1/2$ square lattice Heisenberg AF model with $J/k_{\rm B}=98$\,K.}\label{susceptibility}
\end{figure}

Figure \ref{susceptibility} shows the temperature dependence of the magnetic susceptibility $\chi$ for powder samples of $\lambda$-(BEDT-TTF)$_2$GaCl$_4$. The core diamagnetic contribution of $-4.67\times 10^{-4}$ emu/mol f.u. was already subtracted. The value of $\chi$ at room temperature is $7\times 10^{-4}$ emu/mol f.u., which is greater than the value $4.4\times 10^{-4}$ emu/mol f.u. of $\kappa$-(BEDT-TTF)$_2X$ salts \cite{Kawamoto1995}, suggesting that the electronic state of $\lambda$-(BEDT-TTF)$_2$GaCl$_4$ is more localized than that of the $\kappa$-type salts. $\chi$ increases with decreasing temperature and exhibits a broad hump structure that originates from the low-dimensional magnetic interaction network. We could explain the behavior of the $\chi$ by using the two-dimensional $S=1/2$ square lattice Heisenberg AF model \cite{Lines1970} with exchange interaction $J/k_{\rm B}=98$ K, as shown by the red curve in Fig. \ref{susceptibility}. The Curie constant $C$ was estimated to be 0.45 emu$\cdot$K/mol, indicating that the effective magnetic spin moment ($\mu_{\rm eff}$) is approximately $1 \, \mu_{\rm B}$, which is similar to that of a typical organic Mott insulator $\beta'$-(BEDT-TTF)$_2$ICl$_2$, whose moment is 1 $\mu_{\rm B}$ per dimer \cite{Eto2010b}. Below 13 K, $\chi$ shows an increase. The increase may be due to the canting antiferromagnetism or impurities.

\begin{figure}[htbp]
  \centering\includegraphics[width=9cm,pagebox=cropbox,clip]{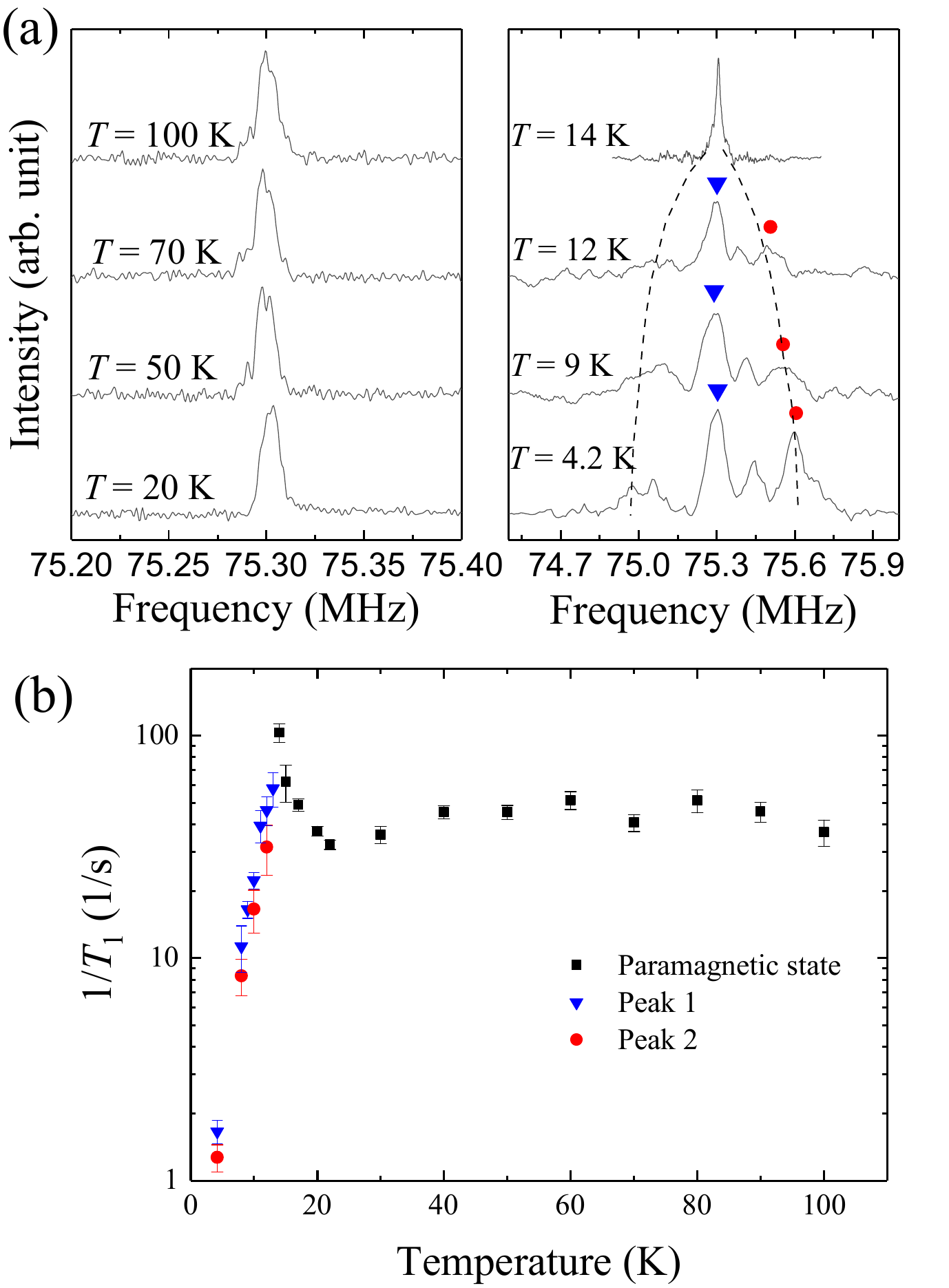}
  \caption{(Color online) (a) $^{13}$C NMR spectra at several temperatures. The dashed lines are guides to the eye. (b) 1/$T_1$ as a function of temperature. Peak 1 and peak 2 correspond to the blue inverted triangle and the red circle in the NMR spectra, respectively.}\label{NMR}
\end{figure}

To clarify the detailed magnetic properties, we performed $^{13}$C NMR measurements. Figure \ref{NMR}(a) shows the temperature dependence of the $^{13}$C NMR spectra from 100 to 4.2 K under a magnetic field of 7 T applied in the [110] direction. In $\lambda$-(BEDT-TTF)$_2$GaCl$_4$, there are two crystallographically non-equivalent BEDT-TTF molecules in a unit cell, and each molecule has two non-equivalent $^{13}$C sites. Therefore, four independent peaks are expected. As confirmed in Fig. \ref{NMR}(a), a broad peak with a fine structure is observed at 100 K, indicating that the four peaks are merged into a broad peak due to almost the same hyperfine coupling constant. Above 20 K, the NMR peak shows no significant temperature dependence. However, the broad peak splits into five discrete peaks below 13 K, and the peak splitting is enhanced as temperature decreases, indicating the emergence of a local field due to AF ordering. The splitting of the NMR peak into a finite number of peaks demonstrates that the local field distributes discretely, indicating that the AF magnetic structure is commensurate.

$1/T_1$ probes the spin fluctuation, which is written as
\begin{eqnarray}
 \frac{1}{T_{1}} \propto \sum _{q} \left( A_{q}A_{-q} \right) \frac{\chi _{q} ''(\omega )}{\omega }.
\end{eqnarray}
Here, $A$ and ${\chi _{q} ''(\omega )}$ are the hyperfine coupling constant and the imaginary part of the dynamic susceptibility at the wave vector $q$, respectively. $\omega$ is the nuclear Larmor frequency. Figure \ref{NMR}(b) shows the temperature dependence of $1/T_1$. It is constant in the temperature range between 100 and 20 K, as expected in a localized spin system well above the N\'eel temperature. $1/T_1$ increases drastically below 20 K due to the critical slowing down and diverges at approximately 13 K,  which coincides with the temperatures where the anomaly in the magnetic susceptibility and the NMR peak splitting are observed. Below 13 K, the temperature dependences of $1/T_1$ of both peak 1 and peak 2, indicated by the blue inverted triangle and the red circle, respectively, in Fig. \ref{NMR}(a), exhibit thermal activation behavior.

\begin{figure}[htbp]
  \includegraphics[width=8.5cm,pagebox=cropbox,clip]{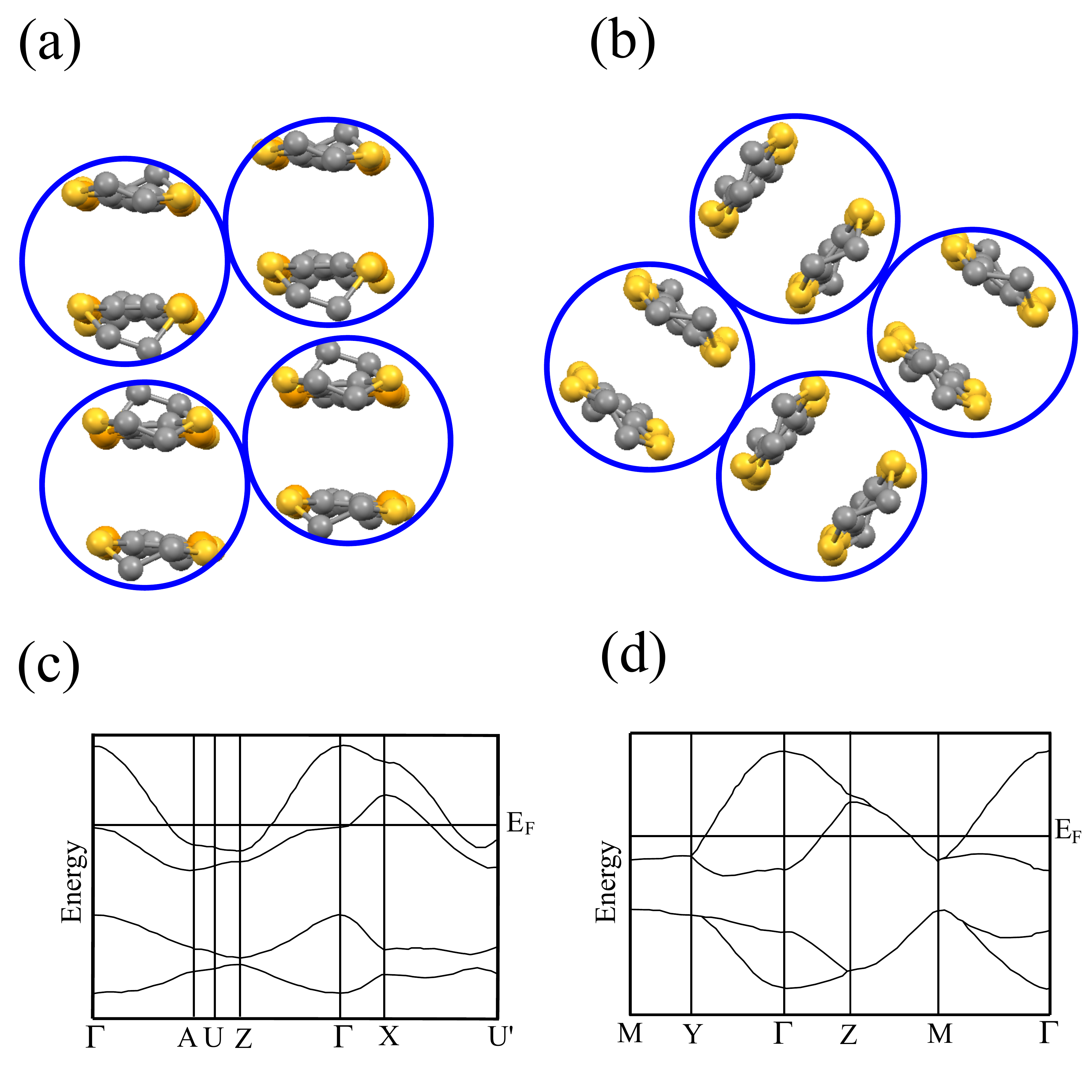}
  \caption{(Color online) Dimer arrangements of (a) $\lambda$-type and (b) $\kappa$-type salts. Circles represent dimers. Band structures of (c) $\lambda$-type \cite{Mori2001} and (d) $\kappa$-type \cite{Oshima1988} salts.}\label{band}
\end{figure}

The commensurate AF structure can be explained by the dimer-Mott insulating picture. Figures \ref{band}(a) and \ref{band}(b) show the donor arrangements of the $\lambda$- and $\kappa$-type salts, respectively. Although the arrangements of the dimers are different from each other, the overlap characteristics of two BEDT-TTF molecules in a dimer are almost the same. The interplanar distance between two BEDT-TTF molecules forming a dimer is 3.769 \AA \cite{Tanaka1999} for $\lambda$-type salts and 3.45--3.53 \AA for $\kappa$-type salts \cite{Saito2015}, and the intradimer transfer integral of $\lambda$-(BEDT-TTF)$_2$GaCl$_4$ is more than twice as large as the other transfer integrals \cite{Kobayashi1996,Mori2001}, suggesting that $\lambda$-type salts have a dimerized structure as in $\kappa$-type salts.
The extended Hu\"ckel tight-binding band calculation showed the upper two and lower two bands with the large energy gap due to dimerization in $\lambda$-type salts, as shown in Fig. \ref{band}(c) \cite{Mori2001}, which is similar to the case of $\kappa$-type salts in Fig. \ref{band}(d) \cite{Oshima1988}. Because GaCl$_4^{-}$ is a monovalent anion and the formal charge of a donor molecule is +0.5, the lower two bands are fully filled and the upper two bands are half-filled. Hence, the electronic system is regarded as a half-filled system, called the dimer-Mott system, as in $\kappa$-type salts. Hartree--Fock calculations suggested that the AF dimer-Mott state can appear in $\lambda$-type salts \cite{Seo1997}. In a dimer-Mott system, one spin is localized on one dimer unit due to on-site Coulomb repulsion, and the AF spin fluctuation from the exchange interaction $J$ is expected.

We could evaluate $J$ from $1/T_1$, using Moriya's expression \cite{Moriya1956}. The $1/T_1$ of antiferromagnets at a sufficiently high temperature is expressed as
\begin{equation}
 \frac{1}{T_{1}}=\frac{\sqrt{2\pi} g^2 \gamma_{\rm N}^2 A_{\rm hf}^2S(S+1)}{3\omega_{ex}},
\end{equation}
where
\begin{equation}
\omega_{ex}^2=\frac{2zJ^2S(S+1)}{3\hbar^2}.
\end{equation}
Here, $g$ is the $g$ value of the electron, $\gamma_{\rm N}$ is the gyromagnetic ratio of the $^{13}$C nuclei, and $z$ is the number of nearest-neighbor spins. $A_{\rm hf}$ is the hyperfine coupling constant. We assumed that $z$ is 4 for the square lattice system, and $A_{\rm hf}$ is 2.9 kOe/$\mu_{\rm B}$ from the results in $\beta'$-type salts, which have the same overlapping mode of BEDT-TTF in a dimer \cite{Eto2010b,Saito2015}. $1/T_1$ is 45 1/s and constant in the high-temperature region. From the $1/T_1$ value, $J/k_{\rm B}$ was estimated to be 112 K, which is in good agreement with the $J/k_{\rm B}\simeq 98$ K value evaluated from the magnetic susceptibility measurement. This result also supports the electronic state of this salt described by the dimer-Mott picture, according to which the localized spins interact with $J$, and the ground state of $\lambda$-(BEDT-TTF)$_2$GaCl$_4$ is the AF ordered state rather than the SDW state due to the nesting of the Fermi surface observed in the TMTSF [TMTSF = tetramethyltetraselenafulvalene] system \cite{Takahashi1986,Nagata2013}.

Because our result revealed that the ground state of $\lambda$-(BEDT-TTF)$_2$GaCl$_4$ is the AF state, the insulating ground state adjacent to the SC phase in the phase diagram of $\lambda$-type salts is expected to be the AF insulating phase. However, magnetic ordering of $\lambda$-(us-BEDT-STF)$_2$GaCl$_4$, which is situated between $\lambda$-(BEDT-TTF)$_2$GaCl$_4$ and $\lambda$-(BETS)$_2$GaCl$_4$ as shown in Fig. \ref{phasediagram}(a), is not reported \cite{Minamidate2015}. Therefore, it is interesting to investigate whether long-range ordering really occurs in $\lambda$-(us-BEDT-STF)$_2$GaCl$_4$. For a more detailed discussion on the connection between the insulating and SC phases, further investigation is required.

Finally, we mention the magnetic nature of $\lambda$-(BETS)$_2$FeCl$_4$, which is isostructural with $\lambda$-(BETS)$_2$GaCl$_4$ and has magnetic anions. It is a coexistent system of $3d$ spins in the FeCl$_4$ layers and $\pi$ electrons in the BETS molecule layers.
It exhibits an AF ordering in zero magnetic field accompanied by a metal--insulator transition at 8.3 K\cite{Kobayashi1996}.
The $3d$ spins in $\lambda$-(BETS)$_2$FeCl$_4$ were considered to play the primary role in the AF ordering, as in a similar system of $\kappa$-(BETS)$_2$Fe$X_4$ ($X$ = Cl, Br) \cite{Otsuka2001,Fukuoka2016}.
However, Akiba \textit{et al}. performed specific heat measurements for $\lambda$-(BETS)$_2$FeCl$_4$ and revealed that the $3d$ spin system shows a Schottky-like anomaly in the magnetic heat capacity below $T_{\rm N}$, and they suggested that the localized $\pi$ electron system undergoes an AF ordering at $T_{\rm N}$, while the $3d$ spin system maintains a paramagnetic state under the effective magnetic field arising from the $\pi{\rm -}d$ interaction. \cite{Akiba2012}. However, the AF ordering state had not been found in $\lambda$-type systems with nonmagnetic anions. The observed AF phase in $\lambda$-(BEDT-TTF)$_2$GaCl$_4$ may yield information on the nature of the AF ordering in $\lambda$-(BETS)$_2$FeCl$_4$. Our result revealed that $\lambda$-type salts are a dimer-Mott system, and there is an AF Mott insulating phase adjacent to the SC phase. Hence, there is a possibility that in the case of $\lambda$-(BETS)$_2$FeCl$_4$, which has localized $3d$ spins, the $\pi$ electron system undergoes the AF ordered state adjacent to the SC phase.

In this study, we performed $^{13}$C NMR and magnetic susceptibility measurements to investigate the magnetic nature of the insulating ground state of $\lambda$-(BEDT-TTF)$_2$GaCl$_4$. We observed NMR spectrum splitting and the divergence of $1/T_1$ at 13 K, which indicate that an AF ordering occurs. This is the evidence that the magnetic ground state of $\lambda$-(BEDT-TTF)$_2$GaCl$_4$ is the AF ordered state. From the discrete NMR peak splitting below $T_{\rm N}$, we revealed that the AF structure is commensurate. The commensurate AF ordering could be understood by the dimer-Mott insulating picture suggested by the band calculation. From the temperature dependence of the magnetic susceptibility, the exchange interaction ($J/k_{\rm B}$) is estimated to be approximately 98 K, which is in good agreement with that estimated from $1/T_1$. Our results suggest that the insulating phase adjacent to the SC phase in $\lambda$-type salts is the AF dimer-Mott insulating phase.

\section*{Acknowledgment}
This work was supported by JSPS KAKENHI Grant Number JP16K0542706.


\end{document}